\documentclass[runningheads]{svmult}

\usepackage{makeidx}   
\usepackage{graphicx}  
\usepackage{subeqnar}  
\usepackage{multicol}  
\usepackage{physprbb}  
\makeindex             

\begin{document}
\title*{On the Emergence of Nonextensivity at the Edge of Quantum Chaos}
%
%
%
%
\titlerunning{On the Emergence of Nonextensivity at the Edge of Quantum Chaos}
%
\author{Yaakov S. Weinstein\inst{1}
\and Constantino Tsallis\inst{2}
\and Seth Lloyd\inst{3} \thanks{yaakov@mit.edu, tsallis@cbpf.br, slloyd@mit.edu}}
\authorrunning{Yaakov S. Weinstein et al.}
%
%
\institute{Massachusetts Institute of Technology, 
Department of Nuclear Engineering
Cambridge MA 02319, USA
\and Centro Brasileiro de Pesquisas Fisicas, Xavier Sigaud 150, 22290-180,
Rio de Janeiro-RJ, Brazil 
\and d'Arbeloff Laboratory for Information Systems and Technology, 
Massachusetts Institute of Technology, Department of Mechanical Engineering, 
Cambridge, Massachusetts 02139, USA \\}

\maketitle              

\begin{abstract}
We explore the border between regular and chaotic quantum dynamics, 
characterized by a power law decrease in the overlap between 
a state evolved under chaotic dynamics and the same state evolved under a 
slightly perturbed dynamics. This region corresponds 
to the edge of chaos for the classical map from which the quantum 
chaotic dynamics is derived and can be characterized via nonextensive 
entropy concepts. 
\end{abstract}

\section{Introduction}
Classical chaotic dynamics is identified by extreme sensitivity to 
initial conditions. Under chaotic dynamics, 
two arbitrarily close points in phase space 
diverge at an exponential rate, quantified 
by the Lyapunov exponent \cite{L+L}. Non-chaotic dynamics does not 
show this extreme dependence on initial conditions, hence, the 
Lyapunov exponent is equal to zero. However, it has been conjectured, that,
though the Lyapunov exponent may vanish, the system may have a 
positive generalized Lyapunov coefficient \cite{C3} 
describing power-law, rather than exponential, divergence of 
classical trajectories. This is the case, at the border between chaotic and 
non-chaotic dynamics (the `edge of chaos') where the Lyapunov exponent goes to 
zero, but there remains a positive generalized Lyapunov coefficient.

This paper identifies a characteristic signature for the edge of quantum chaos.
Quantum states maintain a constant overlap fidelity (heretofore referred to as
the overlap or fidelity) , or distance, under all 
quantum dynamics, regular and chaotic. One way to characterize quantum chaos
is to compare the evolution of an initially chosen state under the chaotic
dynamics with the same state evolved under a perturbed dynamics \cite{P1,P2,SC}. When the initial state is in a regular region of a 
mixed phase space system, a system whose phase space has regular and chaotic
regions, the overlap remains close to one. When the initial state is in a 
chaotic region, the overlap decay is exponential. This paper explores the 
edge of quantum chaos, a region of polynomial overlap decay \cite{YSW1}. 

This paper is structured as follows, we first give a short review of the 
Lyapunov description of chaotic dynamics and the lack of correspondence between
this description and quantum dynamics. We then discuss suggested 
characteristics of quantum chaos (known as signatures of quantum chaos), 
concentrating on the overlap decay first introduced by Peres \cite{P1,P2} . Overlap decay proves to be a useful 
signature of quantum chaos from which to explore the 
border between regular and chaotic quantum dynamics. Next, we will review 
the nonextensive entropy form introduced in \cite{C1} and discuss its 
relevance to the edge of chaos phenomenon. Finally, we locate the 
`edge of quantum chaos' in a mixed phase space system and, using the 
nonextensive entropy formalism, show how the 
overlap decay at the edge of quantum chaos depends on perturbation strength 
and Hilbert space dimension. 

\section{Classical Chaos}

The Lyapunov exponent description of classical chaos is as follows \cite{L+L}.
Let $\Delta x$ be the distance between two points on phase space. We define
$\xi = lim_{\Delta x(0) \rightarrow 0}(\frac{\Delta x(t)}{\Delta x(0)})$, 
to describe how far apart two initially arbitrarily close points on phase space
become at some time $t$. Generally, $\xi(t)$ is the solution to the 
differential equation 
\begin{equation}
\frac{d\xi(t)}{dt}=\lambda_1\xi(t), 
\end{equation}
giving the solution
\begin{equation}
\xi(t) = e^{\lambda_1t}
\end{equation}
where $\lambda_1$ is the Lyapunov exponent (the use of the subscript will 
become clear later on). As seen from the above equation
when the Lyapunov exponent is positive two arbitrarily close points on phase 
space diverge at an exponential rate. Thus, the dynamics described by $\xi(t)$ 
is strongly sensitive to initial conditions and we have chaotic dynamics.

\section{Quantum Chaos}

While the Lyapunov exponent description of chaos works well for points on a 
classical phase space it cannot hold true for quantum mechanical states or
wavefunctions. A measure of distance between quantum 
wavefunctions is the overlap 
\begin{equation}
O_i = \langle\Psi|\Phi\rangle.
\end{equation}
However, the overlap between two quantum wavefunctions 
remains unchanged under unitary evolution governed by the 
linear Schr\"{o}dinger equation. This is seen from
\begin{equation}
O(n) = \langle\Psi (U^n)^{\dag} U^n |\Phi\rangle = O_i,
\end{equation}
where $U$ is the unitary system evolution. Hence, the distance between
two arbitrarily close quantum mechanical wavefunctions, 
like the distance between two Liouville probability densities,  
does not diverge and cannot be described by the Lyapunov exponent picture.
This seeming lack of correspondence has led to the study of `quantum chaos,'
the search for characteristics of quantum dynamics that manifest themselves
as chaotic in the classical realm \cite{B1,B2,H1,BGS}. 

Many such characteristics, quantum signatures of chaos, have been 
detailed in the literature and tested for quantum analogs of classically 
chaotic systems. These signatures can be divided into two broad 
categories, static signatures and dynamic signatures. Static signatures look 
at characteristics of the Hamiltonian or unitary operator governing the 
system. The conjecture is that the evolution operator of quantum chaotic 
systems have statistical properties similar to those of random matrices. 
Hence, quantum analogs 
of classically chaotic systems show level repulsion, that is, if  
the energy eigenvalues of the system are ordered, the difference between 
nearest neighbors would result in a histogram with a Wigner-Dyson 
distribution \cite{BGS} and not 
of a Poisonnian distribution expected for regular systems. 
In addition, the magnitude of the elements of the eigenvectors of 
quantum chaotic operators follow $\chi^2_\nu$ 
distributions\cite{Zyc} from appropriate random matrix ensemble.

Dynamic signatures of quantum chaos look at the evolution of a state under
the quantum chaotic operator compared to the same evolution with some 
additional perturbation. An example of a dynamic signature of chaos is 
hypersensitivity to perturbation. For chaotic systems (both classical 
and quantum), the amount of information necessary to 
track the state of a system when the dynamics is interrupted by an
unknown perturbation grows at an exponential rate with increasing time 
\cite{SC}. Other signatures of quantum chaos look at the entropy growth
of chaotic systems versus regular systems \cite{ZP1}.

\section{Overlap Decay}

Overlap decay as a signature of quantum chaos was first introduced by Peres 
\cite{P1,P2,P3} as a quantum analog of initial state 
sensitivity. Rather
than look at two slightly different states and see how they evolve under 
a certain dynamics, Peres suggested looking at one state, $|\psi_i\rangle$,
and see how it evolves under under two slightly different dynamics, 
an unperturbed Hamiltonian $H$, and the same Hamiltonian with a small 
perturbation $H+\delta V$, where $\delta$ is the perturbation strength. 
The overlap at time $t$ is 
\begin{equation}
O(t) = |\langle\psi_u(t)|\psi_p(t)\rangle|.
\end{equation}
where $|\psi_u(t)\rangle = e^{-iHt}|\psi_i\rangle$ and 
$|\psi_p(t)\rangle = e^{-i(H+\delta V)t}|\psi_i\rangle$ are the unperturbed 
and perturbed states, respectively. The initial behavior of the overlap 
shows different 
behavior depending on whether or not $H$ is chaotic. Recently, the study of 
overlap decay has seen a revival of interest which has served to 
identify several regimes of overlap decay behavior based on whether the system
is chaotic or regular, the type and strength of perturbation, and the type of 
initial state. We note that many of the works cited 
use $O^2$ as the fidelity. Here, we follow \cite{Prosen} and simply 
use the overlap, $O$.

For quantum chaotic systems, quantum versions of classically chaotic maps or
random matrix models, there are several regions of behavior based on the 
strength of the perturbation. Perturbation strength is characterized by 
the size of a typical off diagonal element, $\sigma$, of the perturbation 
operator when the perturbation operator is written in the ordered eigenbasis 
of the unperturbed dynamics, $H$. If $\sigma$ is less than the average 
level spacing of the unperturbed system, $\Delta$, the perturbation is weak.
If $\sigma > \Delta$, the perturbation is in the Fermi Golden Rule (FGR) 
regime \cite{Jacq,Cerruti}. 
The average level spacing is equal to $2\pi/N$ where $N$
is the dimension of the system Hilbert space. A typical off diagonal  
element of the perturbation operator in the ordered eigenbasis of the 
unperturbed system (where the unperturbed system is assumed to have 
eigenvector statistics of a random matrix see \cite{J}) 
is equal to $\sqrt{\delta^2\overline{V_{mn}^2}}$, where
$\overline{V_{mn}^2}$ is the second moment of the matrix elements of the 
perturbation Hamiltonian. 
 
For short enough time, the overlap decay is quadratic for any perturbation 
strength \cite{P1}. After this time, weak perturbations
lead to a Gaussian overlap decay as expected from first order perturbation
theory \cite{P3}. Perturbations in the FGR regime 
lead to an exponential decay of overlap
whose rate, $\Gamma$, increases with perturbation strength. For many systems,  
$\Gamma$ increases as the perturbation strength squared \cite{Jacq}, 
however, some systems do not show this exact dependence \cite{W1}. 
For systems with a classical analog, initial coherent states, and 
classical perturbations, $\Gamma$ will increase with perturbation strength
until the decay rate reaches a value given by the Lyapunov exponent of the 
analog classical system \cite{Jacq,Jala,Cas}. This is a wonderful 
example of a dynamical property of the classical system emerging in quantum 
mechanics. The increase $\Gamma$ may also saturate at the bandwidth of 
$H$ \cite{Jacq}. 

For initial states that are coherent or random states, the Gaussian or 
exponential behavior continues until the fidelity reaches a saturation point 
at $\simeq 1/N$. For initial states that are system eigenstates, there is no 
decay in the weak perturbation regime, and in the FGR regime there is 
an exponential decay of overlap which saturates at $\simeq 1/\Gamma$
\cite{W1,YSW2}.

Regular, non-chaotic, systems have a Gaussian decay in the FGR regime
\cite{Prosen}. The
Gaussian decay is faster then the exponential decay of chaotic states. This
result has been explained using correlation functions and may be 
understood as follows: a perturbation to a chaotic system is quickly spread
out to the entire Hilbert space of the system. Repetition of the same 
perturbation does not lead to a compounded error (the correlation time is 
short). Regular systems, however, do not have this mechanism to spread out the 
perturbation. The same perturbation compounds the error (there is a long
correlation time) leading to a faster fidelity decay. For some more complex
regular systems, a power-law decay 
has also been observed \cite{Been}.

\begin{figure}
\begin{center}
\includegraphics[width=.7\textwidth]{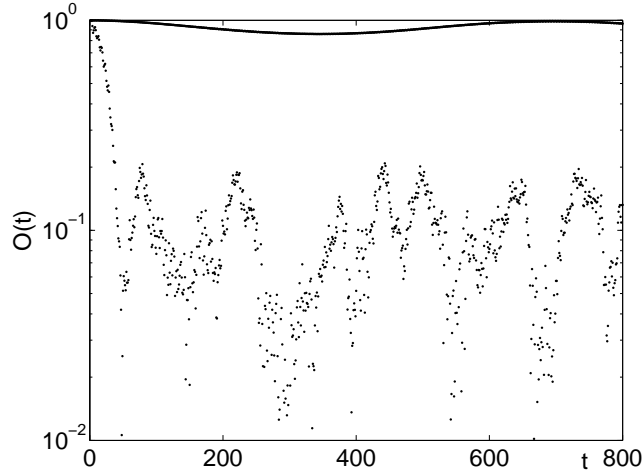}
\end{center}
\caption{
Overlap decay of coherent states in the regular and chaotic zone of the 
$J = 480$ QKT with $k = 3$ and $\delta = .005$ (in the FGR regime). 
The overlap of the 
coherent state centered at the fixed point of the classical kicked top 
is practically an eigenstate of the QKT and oscillates close to unity. The 
overlap of the coherent state in the chaotic zone decays first quadratically 
and then exponentially.   
}
\label{RegChaos}
\end{figure}

Here we study a mixed system, a system with both chaotic and regular regimes.
Coherent states within the regular regime are practically 
eigenstates of the system and the overlap of these states 
oscillates close to unity \cite{P2,Prosen}, as shown in figure
\ref{RegChaos}. 
Coherent states in the chaotic regime behave like other chaotic systems,
they show exponential overlap
decay in the FGR regime and Gaussian overlap decay for weak perturbations. 
We explore the border between the regular and chaotic regions and show that, 
in both the FGR and weak perturbation regimes, coherent states near this 
border have a polynomial overlap decay \cite{YSW1}.

\section{Nonextensive Entropy}

The Boltzmann-Gibbs formulation of statistical mechanics is one of the pillars
of modern day physics. However, its applicability, indeed its formulation, is
only for systems with short range interactions. Self-gravitating systems and 
other systems such as those  with long range forces, long range memories 
(non-Markovian), or multifractal structures present significant difficulties 
for the application of the Boltzmann-Gibbs formulation. In 1988, in an 
attempt to understand the nature of some of these systems, one of us 
\cite{C1} proposed the generalization of Boltzmann-Gibbs statistical mechanics on the basis of a generalized non-extensive entropy, $S_q$. 
When $q = 1$, $S_q$ recovers the usual Boltzmann-Gibbs entropy.

The generalized entropy is given as follows:  
\begin{equation}
S_q = k \frac{   1-\sum^W_{i=1}p^q_i  }{q-1}
\end{equation}
where $k$ is a positive constant, $p_i$ is the probability of finding the 
system in microscopic state $i$, and $W$ is the number of possible 
microscopic states of the system; $q$ is the entropic index which 
characterizes the degree of the system nonextensivity. In the limit 
$q \to 1$ we recover the usual Boltzmann entropy 
\begin{equation}
S_1 = -k\sum^W_{i=1}p_i\; lnp_i. 
\end{equation}
To demonstrate how $q$ characterizes the degree of nonextensivity of a system,  
we present the $S_q$ entropy addition rule \cite{C2}. If $A$ and
$B$ are two independent systems such that the probability $p(A+B) = p(A)p(B)$, 
the entropy of the total system $S_q(A+B)$ is given by the following:
\begin{equation}
\frac{S_q(A+B)}{k}=\frac{S_q(A)}{k}+
\frac{S_q(B)}{k}+(1-q)\frac{S_q(A)S_q(B)}{k^2}.
\end{equation}
From the above equation one can see that $q<1$ corresponds to 
superextensivity, the combined system entropy is more then the sum of the two 
independent systems; $q>1$ corresponds to subextensivity, the combined system
entropy is less then the two independent systems. Using this entropy 
to generalize statistical mechanics and thermodynamics has helped to explain 
many natural phenomena in a wide range of fields.  

One application of nonextensive entropy occurs in one dimensional 
dynamical maps. As explained above, when the Lyapunov 
exponent of a system is positive, the system dynamics is strongly 
sensitive to initial conditions and is called chaotic. When the Lyapunov 
exponent goes to zero it has been conjectured \cite{C3} (and proven for the 
logistic map \cite{robledo}) that the distance between two initially 
arbitrarily close points can be described by 
$\frac{d\xi}{dt} = \lambda_{q_{sen}}\xi^{q_{sen}}$ leading to 
$\xi = [1+(1-q_{sen})\lambda_{q_{sen}} t]^{1/(1-q_{sen})}$ ({\it sen} stands 
for sensitivity). This requires the introduction of 
$\lambda_{q_{sen}}$ as a generalized Lyapunov coefficient. 
The Lyapunov coefficient
scales inversely with time as a power law instead of the characteristic 
exponential of a Lyapunov exponent. Thus, there exists a regime, 
$q_{sen}<1, \lambda_1 = 0, \lambda_{q_{sen}}>0$, which is weakly sensitive to 
initial conditions and is characterized by having power law, instead of 
exponential, mixing. This regime is called  the edge of chaos. 

The polynomial overlap decay found for initial states of a mixed system 
near the chaotic border are at the `edge of quantum chaos', 
the border between regular and chaotic quantum dynamics. This region is
the quantum analog of the classical region characterized by 
the generalized Lyapunov coefficient.

\section{The Quantum Kicked Top}

The system studied in this work is the quantum kicked top (QKT) \cite{H2} 
defined by the operator:
\begin{equation}
U_{QKT} = e^{-i\pi J_y/2\hbar}e^{-i\alpha J_z^2/2J\hbar}.
\end{equation}
$J$ is the angular momentum of the top and $\alpha$ is the `kick' 
strength. The representation is such that $J_z$ is diagonal.
The classical version of the kicked top has either regular, mixed, or 
fully chaotic dynamics depending upon the kick strength. 
We work with a QKT of $\alpha = 3$ whose 
classical analog has a mixed phase space, with clearly defined 
regions of chaotic and regular dynamics. The 
perturbed operator is simply a QKT with a stronger kick strength 
$\alpha'$. Hence, the perturbation operator, $V = \delta \pi J_z^2/2J$ 
where $\delta \equiv \alpha' - \alpha$.

When $J$ is even the QKT has three invariant, dynamically independent
 subspaces \cite{P2,H2}. We can write basis functions 
for the invariant subspaces in terms 
of the eigenvectors of $J_z$ which, following Peres \cite{P2}, 
we will denote as $|m\rangle$.
The {\it ee} subspace is even under a $180^\circ$ rotation about $x$ and 
even under a $180^\circ$ rotation about $y$ and has a Hilbert space dimension 
of $N = J/2 +1$. The basis functions of the {\it ee} subspace are $|0\rangle$
and $(|2m\rangle + |-2m\rangle)/\sqrt{2}$, where $m$ ranges from 1 to $J/2$. 
The {\it oo} subspace is even under a $180^\circ$ rotation 
about $x$ and odd under a $180^\circ$ rotation about $y$ and $N = J/2$. Its 
basis functions are $(|2m-1\rangle - |1-2m\rangle)/\sqrt{2}$. The 
{\it oe} subspace is odd under a $180^\circ$ rotation about $x$, $N = J$, and 
has basis functions $(|2m\rangle - |-2m\rangle)/\sqrt{2}$ and 
$(|2m-1\rangle + |1-2m\rangle)/\sqrt{2}$.
All of the numerical simulations in this work were 
done in the {\it oo} subspace of the QKT. This is to say, we construct the 
complete QKT and transform it into a basis such that it is block diagonal
with the dimensions of the three blocks mentioned above. The columns of the 
transformation matrix $T$ to get the QKT operator into block diagonal form 
are the states which form the bases of the invariant subspaces.  We take 
only the block corresponding to the {\it oo} subspace and use that as our map.
The initial angular momentum coherent states are also transformed into 
this basis and, again, only the part of the state corresponding to the 
{\it oo} subspace is used.

\begin{figure}[b]
\begin{center}
\includegraphics[width=.7\textwidth]{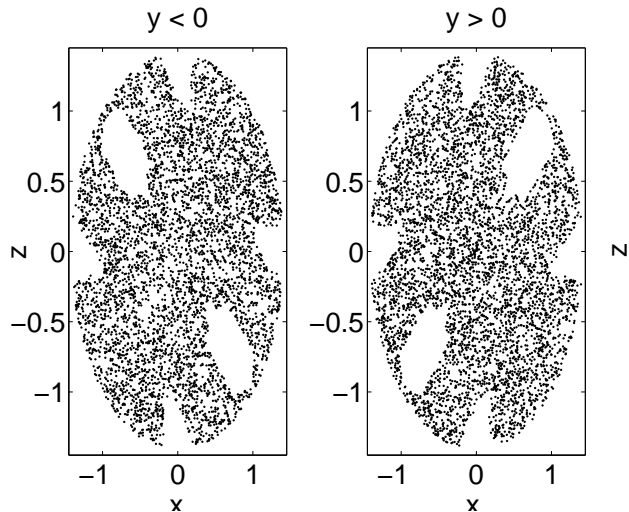}
\end{center}
\caption{10000 points of a chaotic orbit on the phase space of the classical
kicked top with kick strength $\alpha = 3$. The regular regions are clearly 
visible. The spherical phase space is projected onto the $x-z$ plane 
by multiplying the $x$ and $z$ coordinates of each point by $R/r$ where 
$R = \sqrt{2(1-|y|)}$ and $r = \sqrt{(1-y^2)}$ \protect\cite{P2}.}
\label{phase}
\end{figure}

The phase space of the classical kicked top is the unit sphere, 
$x^2+y^2+z^2=1$ and the resulting action of the map is:
\begin{eqnarray}
\begin{array}{cc}
x'= & z\\
y'= & x \;sin(\alpha z)+ y \;cos(\alpha z)\\
z'= & -x\; cos(\alpha z) + y\; sin(\alpha z).
\end{array}
\end{eqnarray}
For $\alpha = 3$ there are two fixed points of order one at the center of the 
regular regions. They are located at
\begin{equation}
x_f = z_f = \pm 0.6294126, \;y_f = 0.4557187. 
\end{equation}
The regular and chaotic regions of the kicked top are clearly seen in 
the classical maps phase space shown  
in the figure \ref{phase}.
Hence, we expect quantum coherent states centered near the classical
periodic point to exhibit significantly different behavior than coherent states
centered in the classically chaotic region of the map. This is indeed seen in 
figure \ref{RegChaos}.

\section{Locating the Edge of Quantum Chaos}

\begin{figure}
\begin{center}
\includegraphics[width=.8\textwidth]{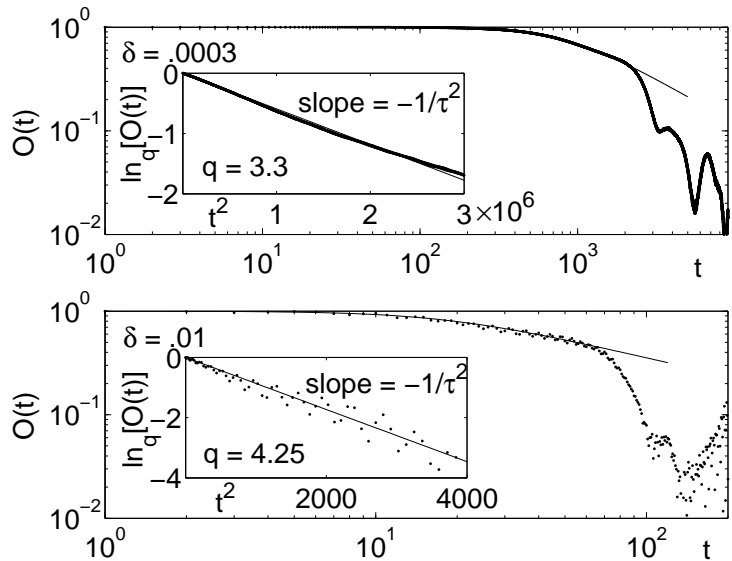}
\end{center}
\caption{Overlap versus time for an initial angular momentum coherent state
located at the border between regular and chaotic zones of the QKT of spin
240 and $\alpha = 3$. This region, the edge of quantum chaos,
shows the expected power law decrease in overlap. The top figure is for a 
perturbation strength in the weak perturbation regime, 
$\delta = .0003$ and the bottom 
figure is for a perturbation strength of $\delta = .01$, within the
FGR regime. On the log-log plot the power law decay region, 
from about 600-2500 in the weak perturbation regime and 20-70 in the 
FGR regime, is linear. We can fit the decrease in overlap with the expression 
$[1+(q_{rel}-1)(t/\tau_{q_{rel}})^2]^{1/(1-q_{rel})}$ where, in the weak 
perturbation  regime, the entropic index $q_{rel} = 3.3$ and 
$\tau_{q_{rel}} = 1300$ and in the FGR regime $q_{rel} = 4.25$ and 
$\tau_{q_{rel}} = 34$. The insets of both figures show 
$\ln_{q_{rel}} O \equiv (O^{1-q_{rel}}-1)/(1-q_{rel})$ versus $t^2$; 
since $\ln_q x$ is the inverse function of 
$e_q^x \equiv [1+(1-q) \; x]^{\frac{1}{1-q}}$, this produces a straight line
with a slope $-1/\tau^2$ (also plotted).
}
\label{edge}
\end{figure}

To locate the edge of quantum chaos we use initial angular momentum 
coherent states keeping $y$ equal to $y_f$ of the positive fixed point for the 
classical kicked top and changing $z$ until a state which has a 
power-law overlap decay is found. The state which gives this behavior 
depends on the angular momentum of the QKT, $J$, but, given a fixed $J$, 
the power law emerges for perturbations 
in both the weak perturbation and FGR regimes. Examples of edge of quantum
chaos behavior in 
both regimes is shown in figure \ref{edge} for a QKT of $J = 240$. 
As seen in the figure, the 
power-law overlap decay is transitory between the quadratic 
and exponential behavior of the overlap decay. This transitory region 
does not appear for chaotic states (as shown in figure \ref{RegChaos}) or 
states close to the fixed point of the map, it is a signature of the 
`edge of quantum chaos.'

The overlap decay for the edge of quantum chaos state is very well fit by the 
solution of the differential equation 
\begin{equation}
dO/d(t^2) = -O^{q_{rel}}/\tau_{q_{rel}}^2.
\end{equation}  
In the above equation, {\it rel} stands for relaxation which is an appropriate 
description of a $q>1$ phenomenon. In classical systems, 
$q_{rel}$ characterizes the relaxation of initial states towards an attractor. 
Although we do not know how to derive this differential equation 
from first principles, the numerical agreement is remarkable 
(see also \cite{borges}). A time-dependent
$q$-exponential expression analogous to the one shown here has recently 
been proved for the edge of chaos and other critical points of the classical 
logistic map \cite{robledo}. 

\begin{table}
\caption{The edge of quantum chaos, the critical perturbation strength, 
$\delta_c$, and $q_{rel}^c$, the $q_{rel}$ for perturbations strengths below 
$\delta_c$, for explored values of $J$. As $J$ increases behavior 
characteristic of the edge of quantum chaos occurs further away
from the fixed point of the classical map. The critical perturbation and 
$q_{rel}^c$ decrease with increased $J$.}
\begin{center}
\renewcommand{\arraystretch}{1.4}
\setlength\tabcolsep{15pt}
\begin{tabular}{@{}llp{1.8cm}l}
\hline\noalign{\smallskip}
J & Edge & $\delta_c$ & $q_{rel}^c$\\
\noalign{\smallskip}
\hline
\noalign{\smallskip}
120 & $z_f - .124$ & $5.39\times 10^{-3}$ & 3.8\\
150 & $z_f - .139$ & $3.86\times 10^{-3}$ & 3.7\\
180 & $z_f - .151$ & $2.94\times 10^{-3}$ & 3.6\\
210 & $z_f - .160$ & $2.33\times 10^{-3}$ & 3.4\\
240 & $z_f - .176$ & $1.91\times 10^{-3}$ & 3.3\\
280 & $z_f - .183$ & $1.51\times 10^{-3}$ & 3.1\\
360 & $z_f - .190$ & $1.04\times 10^{-3}$ & 2.8\\
480 & $z_f - .194$ & $6.74\times 10^{-4}$ & 2.6\\
\noalign{\smallskip}
\hline
\noalign{\smallskip}
\end{tabular}
\end{center}
\label{Tab1}
\end{table}

\begin{figure}
\begin{center}
\includegraphics[width=.7\textwidth]{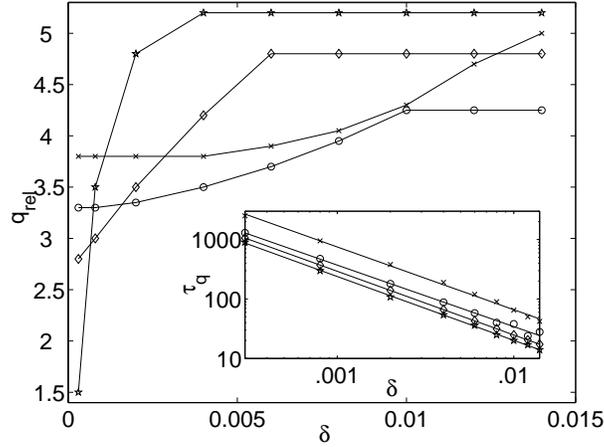}
\end{center}
\caption{Values of $q_{rel}$ and $\tau_q$ (inset) for $J = 120$ (x), 240
(circles), 360 (diamonds) and 480 (stars). $q_{rel}$ remains constant for
perturbation strengths below the critical perturbation and above the 
saturation perturbation. In between $q_{rel}$ increases with a rate 
dependent on $J$. The
values of $q_{rel}^c$, $q_{rel}^s$, $\delta_c$ and $\delta_s$ can be seen in 
the figure. In addition the rate of growth of $q_{rel}$ with increased 
perturbation strength can be seen. The inset shows a loglog plot 
of the value of $\tau_q$ versus $\delta$ for the above values of $J$. The data
can be fit with a lines of slope -1.06, -1.03, -1.07, and -1.08 for $J = 120$,
240, 360 and 480 (top to bottom). 	
}
\label{qt}
\end{figure}

\begin{figure}
\begin{center}
\includegraphics[width=.7\textwidth]{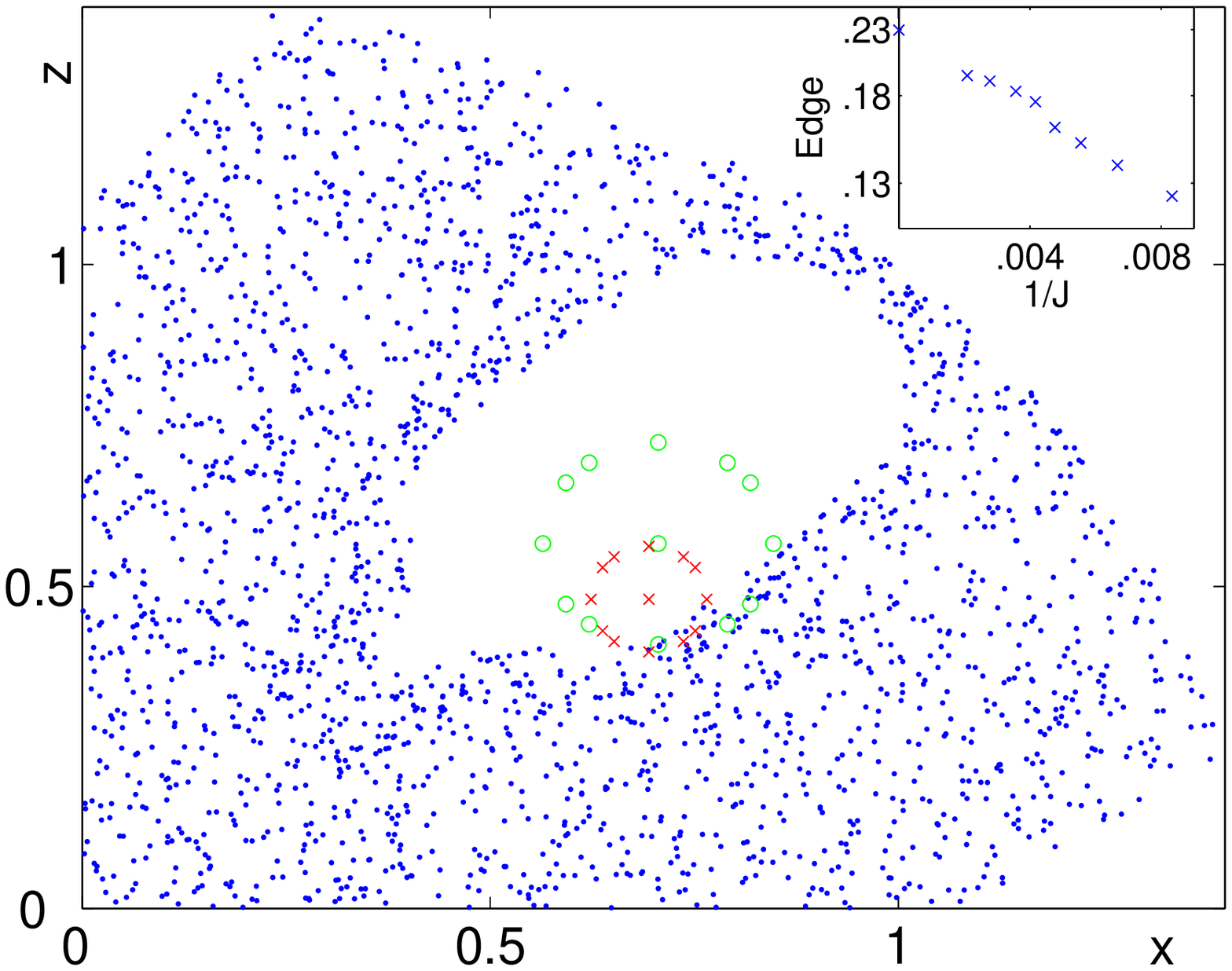}
\end{center}
\caption{Coherent states of $J = 120$ (circles) and $J = 480$ (x) at the 
edge of quantum chaos superimposed on the upper right quarter of the classical
phase space (shown in figure 2). The spherical phase space and ellipsoidal
coherent states are projected onto the $x-z$ plane as described above. 
The $J = 120$ coherent state is significantly larger then the $J = 480$ 
coherent state, causing edge of quantum chaos behavior to appear for
the $J = 120$ wavefunction much closer to the classical periodic point than
edge of chaos behavior for the $J = 480$ wavefunction. The inset is a 
semilog ($y$-axis) plot of $1/J$ versus distance from the
fixed point of the classical map. Note that the point of appearance of  
edge of chaos behavior follows an inverse power law with the size of the 
coherent state. }
\label{edgeW}
\end{figure}

The values of $q_{rel}$ and $\tau_{q_{rel}}$ depend on the perturbation strength, $J$,
and on whether the perturbation strength is in the weak perturbation or in the FGR 
regime. For perturbation strengths below the FGR regime, $\delta$ is less
than some $\delta_c$,
$q_{rel}$ remains constant at a value of $q_{rel}^c$. The transition into the
FGR regime arises when the typical off diagonal 
elements of $V$ are larger than $\Delta$. We can approximate 
$\delta_c \simeq \sqrt{2\pi/N^3}$ \cite{Jacq} where, in the $oo$ subspace of 
the kicked top $N = J/2$. Values for $\delta_c$ are shown in the above table.
In the FGR regime $q_{rel}$ continues to increase with
perturbation strength until a saturation perturbation, $\delta_s$, after 
which $q_{rel} = q_{rel}^s$ remains constant. 
As the top becomes more classical, with increased $J$, there is a decrease in 
$q_{rel}^c$ while in the weak perturbation regime, but
an increase in the rate of increase of $q_{rel}$ while in the FGR regime.
Because of this larger rate of increased $q_{rel}$ in the FGR regime, 
$q_{rel}^s$ increases with increasing $J$. However, we 
see that this is not the case for the $J = 120$ case. For $J = 120$, $q_{rel}$
increases beyond the expected saturation point. This may be due because
the $J = 120$ coherent state is so large that, at stronger 
perturbations, it leaks out of the regular region of the map in more 
than one place due to the odd shape of the regular region (see figure 
\ref{edgeW}).
The value of $\tau_{q_{rel}}$ decreases with perturbation strength and is well 
fit by a line of slope approximately -1 on a log-log plot;  $\tau_{q_{rel}}$ also 
decreases with increasing $J$ at a fixed perturbation strength. The values of 
$q_{rel}$ and $\tau_{q_{rel}}$ for a number of different 
perturbation strengths can be seen in the figure \ref{qt}.

\begin{figure}
\begin{center}
\includegraphics[width=.7\textwidth]{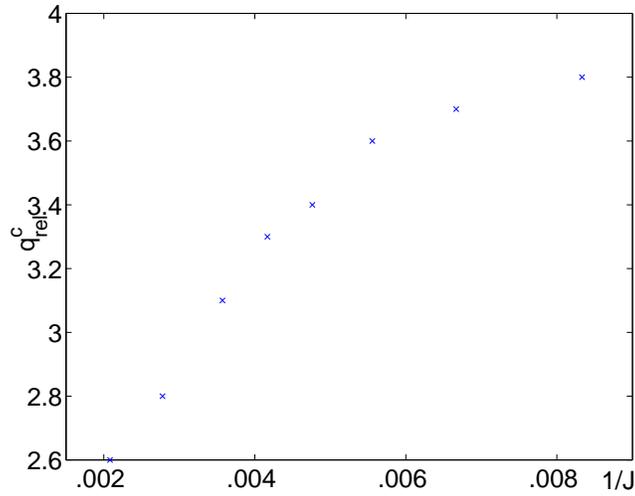}
\end{center}
\caption{Values of $q_{rel}^c$ versus $1/J$. These are determined by exploring
a number of perturbations much less than $\delta_c$.}
\label{qc}
\end{figure}

The location of states 
exhibiting edge of quantum chaos behavior is not the same as the edge of
chaos for the classical kicked top.  This is due to the 
finite size of the coherent states. Though the coherent state may be
centered at a point that is classically regular, some of the state may
'leak out' into the chaotic region of the map. As $J$ increases
the coherent state gets smaller and the quantum edge approaches the classical
one. The location of the edge and the size of the coherent state compared 
to the regular regions of the map are shown in figure \ref{edgeW}. 

In conclusion, we have explored the region at the between chaotic and 
non-chaotic quantum dynamics, the edge of quantum chaos. 
Coherent states located at this border 
exhibit a power-law decrease in overlap as opposed to practically no
decay for coherent states near the periodic point of the classical map
and exponential overlap decay exhibited by fully chaotic quantum dynamics. 
This region is the quantum parallel of the  classical region at the border 
between regular and chaotic classical dynamics where the Lyapunov exponent goes
to zero and the mixing is characterized by the generalized Lyapunov 
coefficient. Further studies of this rich system are certainly welcome. 

One of us (C.T.) acknowledges warm hospitality by H.-T. Elze and the organizers during the interesting meeting.

%

\end{document}